\documentclass{aa}
\usepackage{graphicx}
\usepackage{txfonts}
\usepackage{natbib}
\begin{document}
%
\title{Relationship between non-thermal electron energy spectra and \textit{GOES} classes}

\author{R. Falewicz\inst{1}
          \and
           P. Rudawy\inst{1}
           \and
           M. Siarkowski\inst{2}
}
  \titlerunning{Relationship between non-thermal electron energy spectra and \textit{GOES} classes}

   \offprints{R. Falewicz}

   \institute{Astronomical Institute, University of Wroc{\l}aw,
             51-622 Wroc{\l}aw, ul. Kopernika 11, Poland\\
             \email{falewicz@astro.uni.wroc.pl; rudawy@astro.uni.wroc.pl}
         \and
             Space Research Centre, Polish Academy of Sciences, 51-622 Wroc{\l}aw,
   ul. Kopernika 11, Poland\\
              \email{ms@cbk.pan.wroc.pl}}

   \date{Received xx xx, 2008; accepted xx xx, 2009}

\abstract{} {We investigate the influence of the variations of
energy spectrum of non-thermal electrons on the resulting
\textit{GOES} classes of solar flares.} {Twelve observed flares with
various soft to hard X-ray emission ratios were modeled using
different non-thermal electron energy distributions. Initial values
of the flare physical parameters including geometrical properties
were estimated using observations.} {We found that, for a fixed
total energy of non-thermal electrons in a flare, the resulting
\textit{GOES} class of the flare can be changed significantly by
varying the spectral index and low energy cut-off of the non-thermal
electron distribution. Thus, the \textit{GOES} class of a flare
depends not only on the total non-thermal electrons energy but also
on the electron beam parameters. For example, we were able to
convert a M2.7 class solar flare into a merely C1.4 class one and a
B8.1 class event into a C2.6 class flare. The results of our work
also suggest that the level of correlation between the cumulative
time integral of HXR and SXR fluxes can depend on the considered HXR
energy range.}{}


\maketitle
%

\section{Introduction}

Despite recent progress, the source and acceleration mechanisms of
particles in solar flares are still far from being understood. It is
commonly accepted that, during the flare impulsive phase,
non-thermal electron beams are accelerated in the solar corona and
move along magnetic field lines to the chromosphere where they
deposit their energy. Here, most non-thermal electrons lose their
energy in Coulomb collisions while a tiny part of the electron
energy is converted into hard X-rays (HXR) by bremsstrahlung. The
heated chromospheric plasma evaporates and radiates over a wide
spectral range from hard X-rays or gamma rays to radio emission.
Hard and soft X-ray fluxes emitted by solar flares are generally
related, in a way first described by \citet{Neupert1968}, who found
that the time derivative of the soft X-ray flux approximately
matches the microwave flux during the flare impulsive burst. A
similar effect was also observed for hard X-ray emission
(\citet{Dennis1993}). Since hard X-ray and microwave emissions are
produced by non-thermal electrons and soft X-rays are the thermal
emission of a hot plasma, the Neupert effect suggests that
non-thermal electrons are the direct source of plasma heating.
\citet{Lin1984} were the first to observe hard X-ray emission above
25 keV from microflares using balloon-borne observations. Later
studies using RHESSI observations (\citet{Qiu2004};
\citet{Battaglia2005}; \citet{Hannah2008}) show that time, spatial
and spectral characteristics of microflares are similar to those of
large flares. However, there is no universal, unambiguous
correlation between the released total energy of the flare and the
observed HXR radiation.

Many flares reveal a low SXR emission (as indicated by GOES
emission), but strong emission of the HXR emission above 30 keV
(\citet{mcd99}, \citet{Gburek02}, \citet{Qiu2004}, \citet{Siar06}).
Such events are commonly called "non-correlated" flares. Analytical
estimations made by \citet{mcd99} show that in these flares, only a
small part of the total energy carried by the non-thermal electrons
is transferred to ambient material during the chromospheric
evaporation process. Even so, analysis of the HXR emission of these
flares allows one to estimate the energy flux of non-thermal
electrons and to numerically simulate the energy losses and
hydrodynamic effects of the chromospheric evaporation, thus allowing
investigations of the energy budgets of the solar flares.

\begin{figure*}[ht!]
\centering

\includegraphics[width=5.0cm]{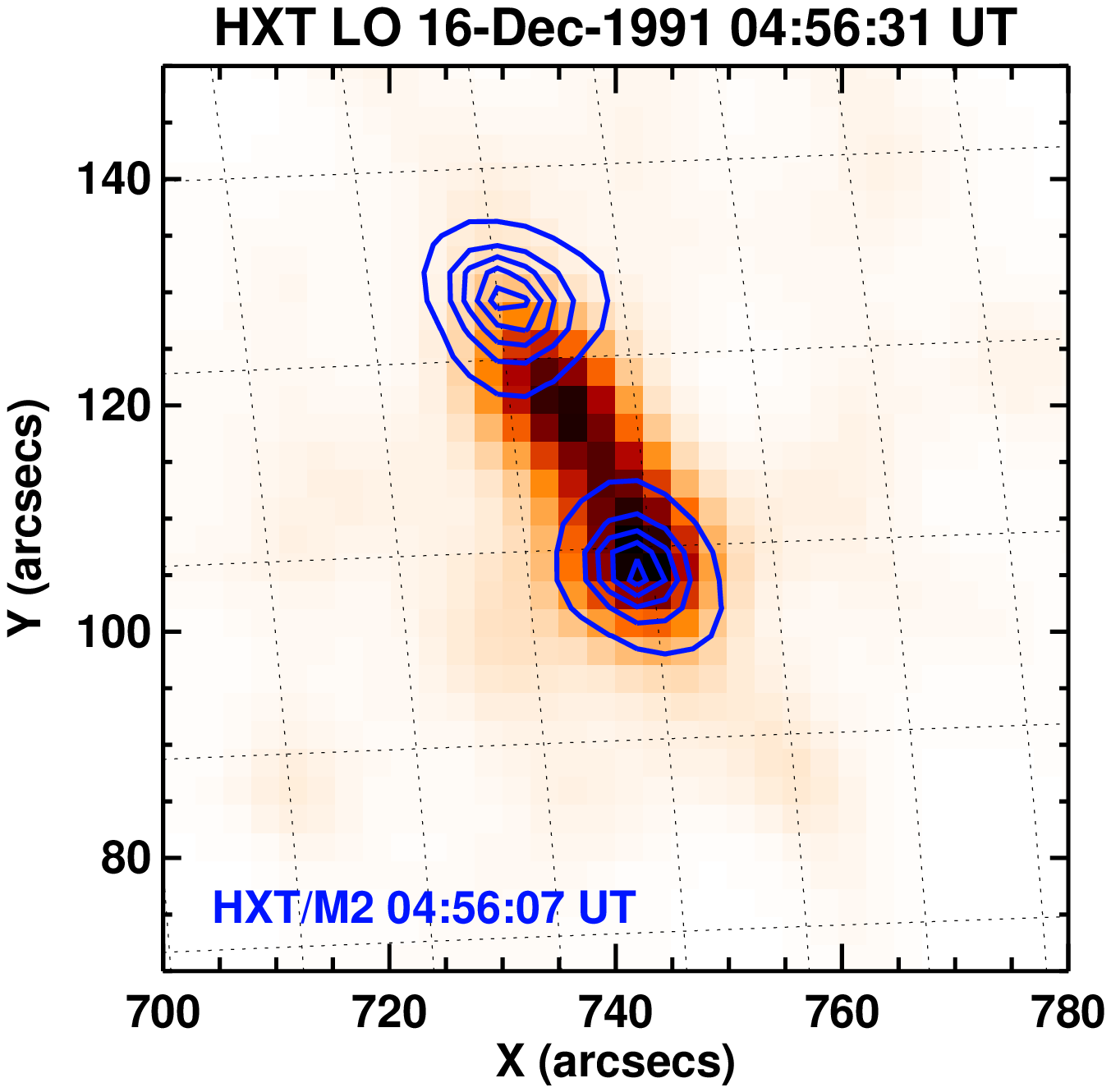}
\includegraphics[width=5.0cm]{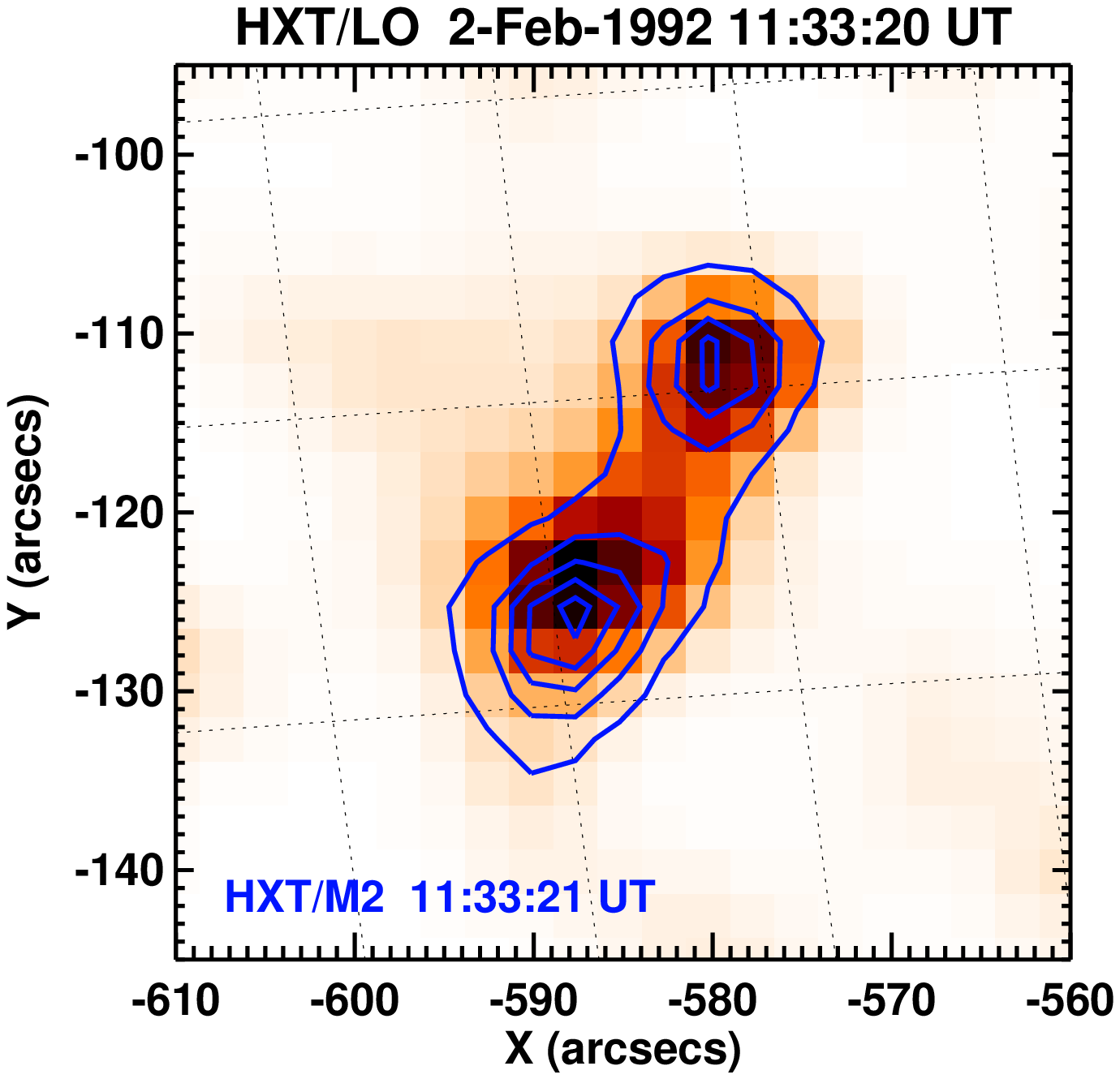}
\includegraphics[width=5.0cm]{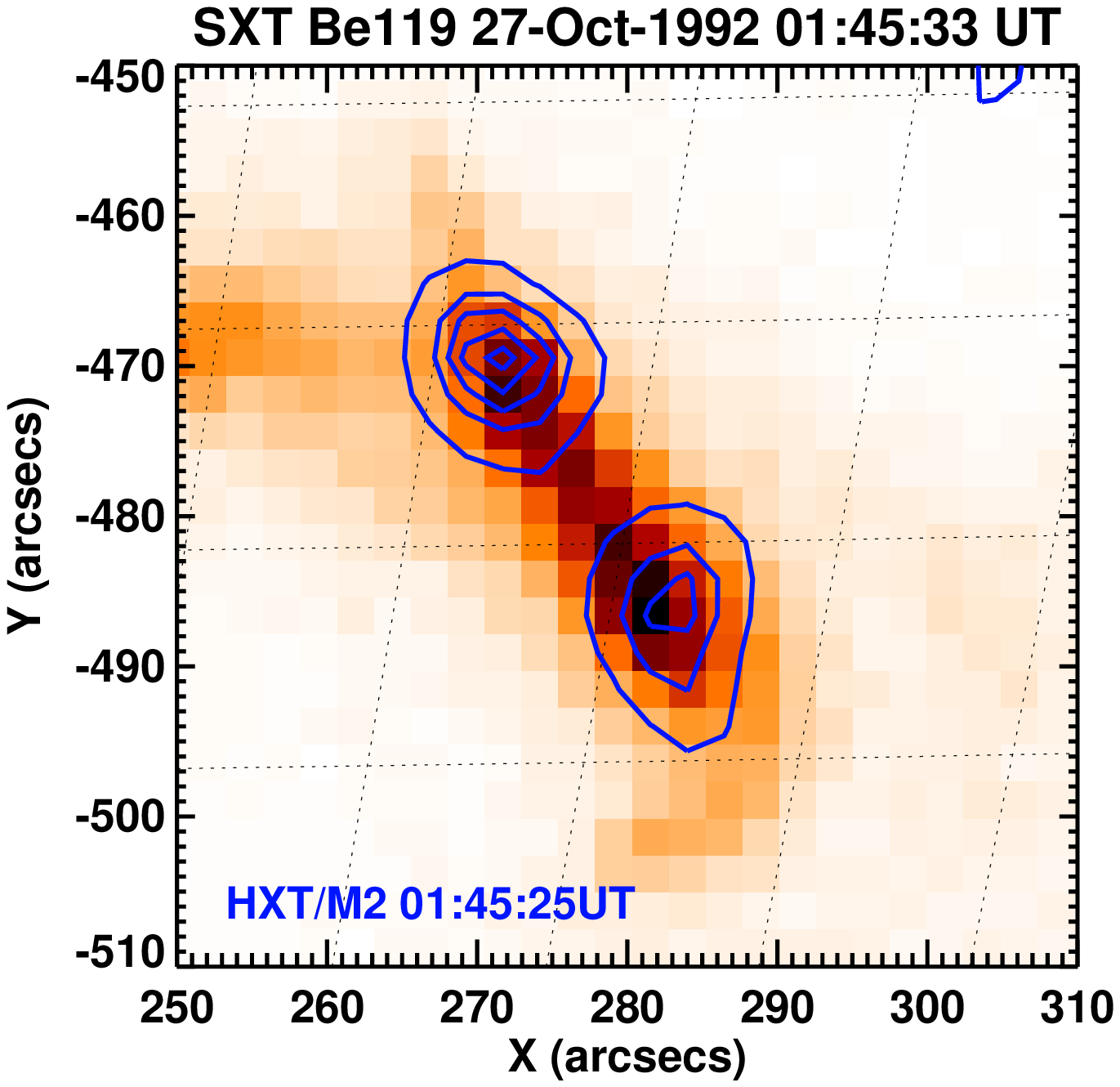}

\includegraphics[width=5.0cm]{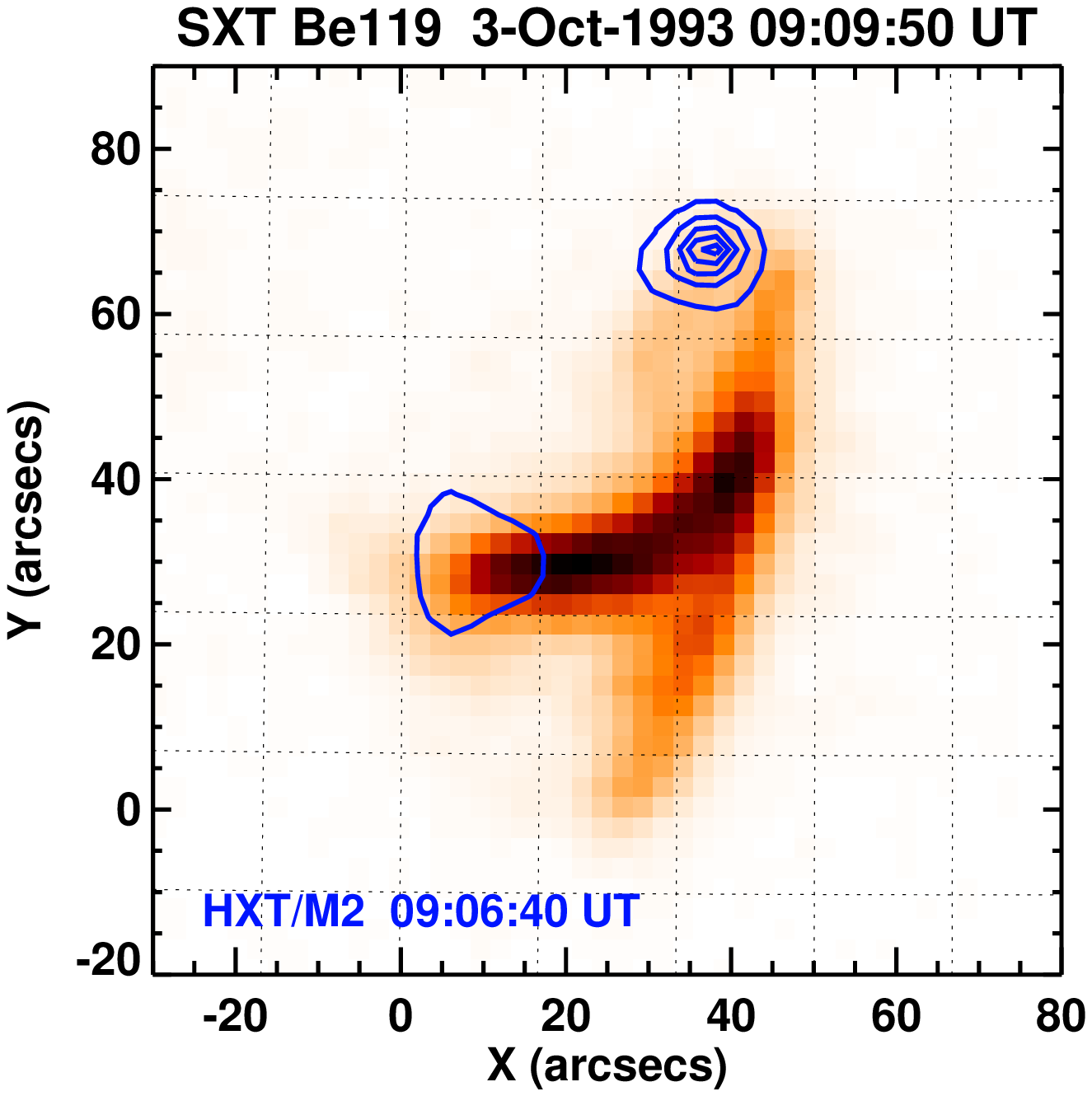}
\includegraphics[width=5.0cm]{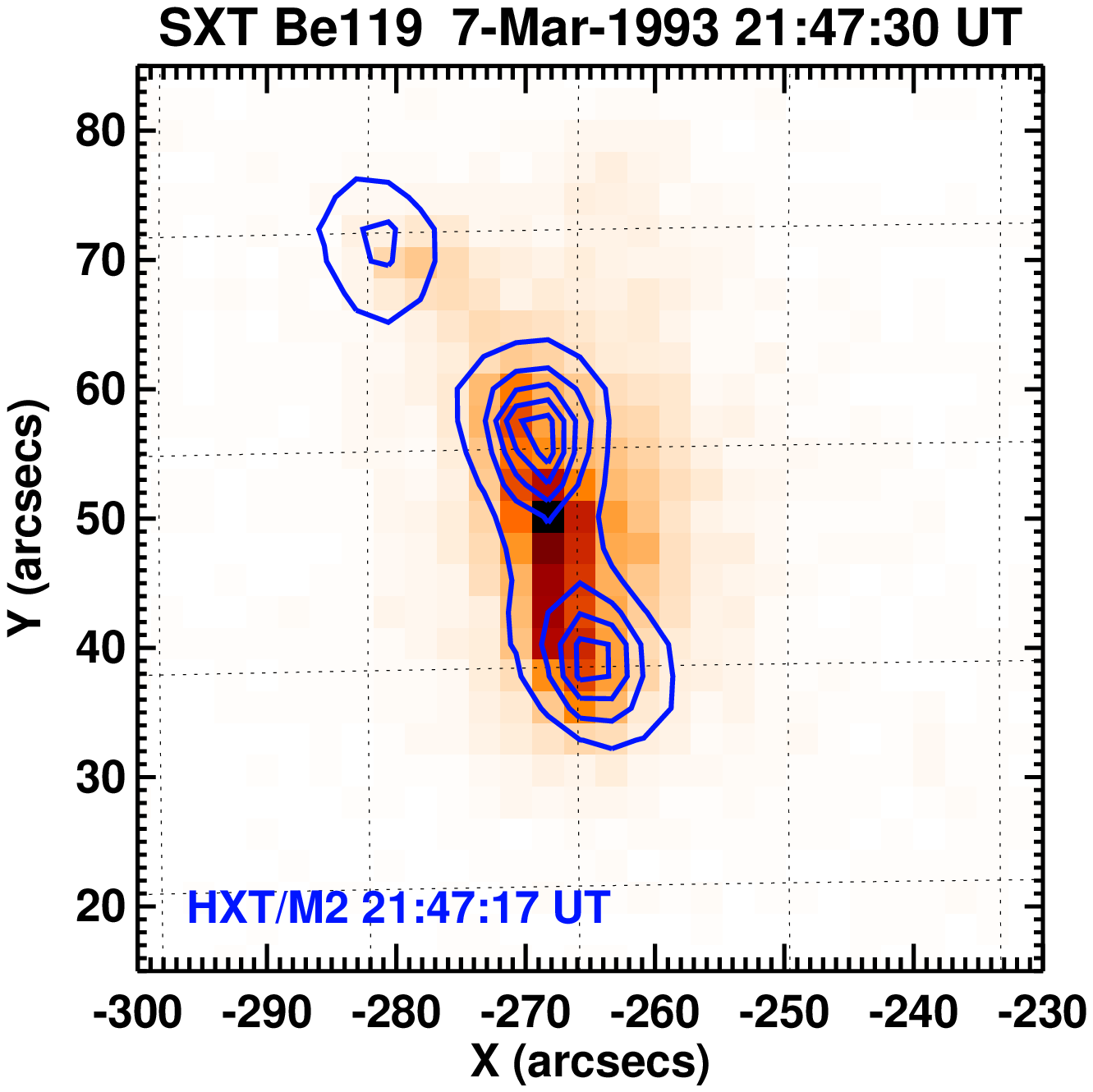}
\includegraphics[width=5.0cm]{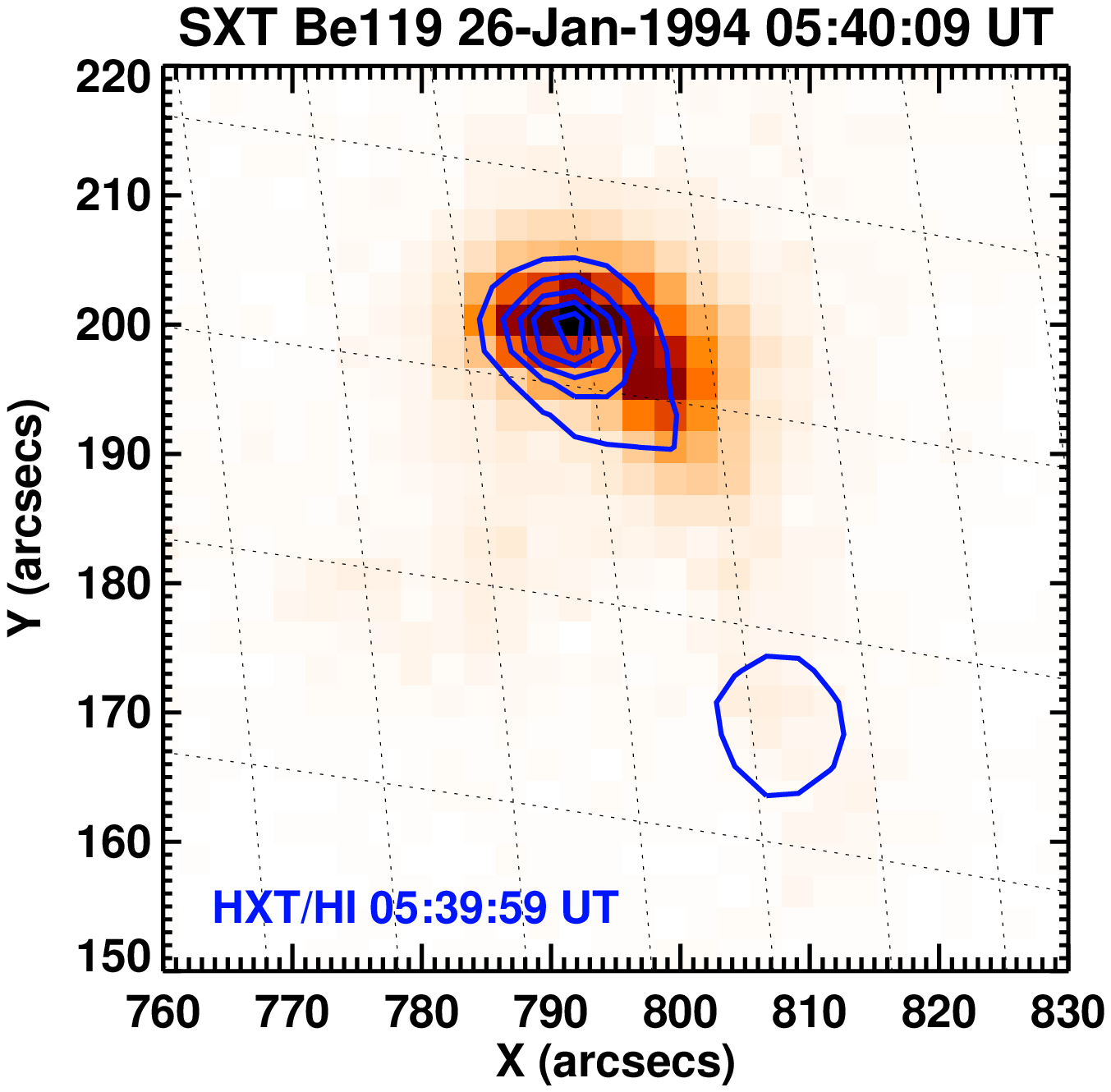}

\includegraphics[width=5.0cm]{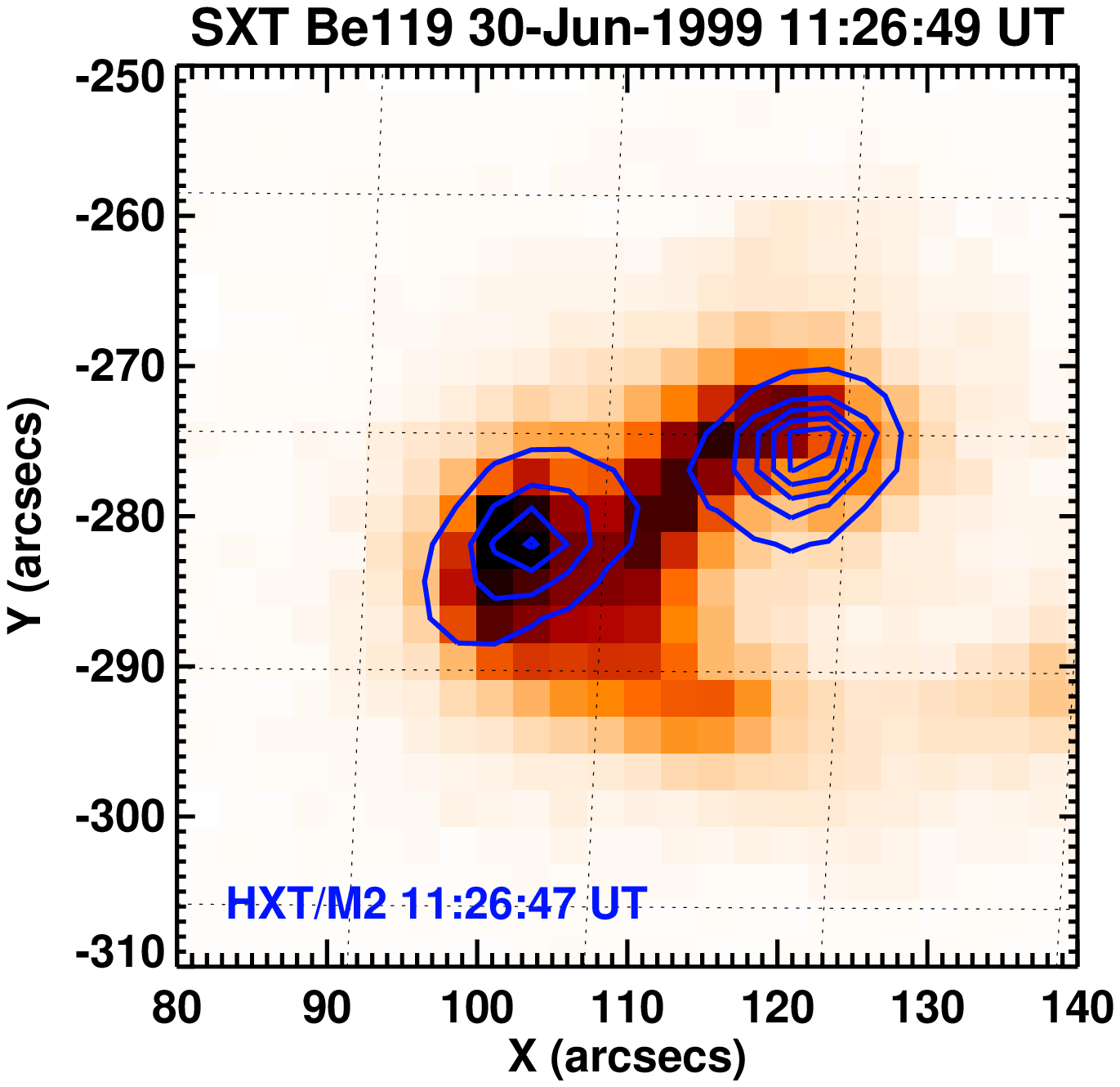}
\includegraphics[width=5.0cm]{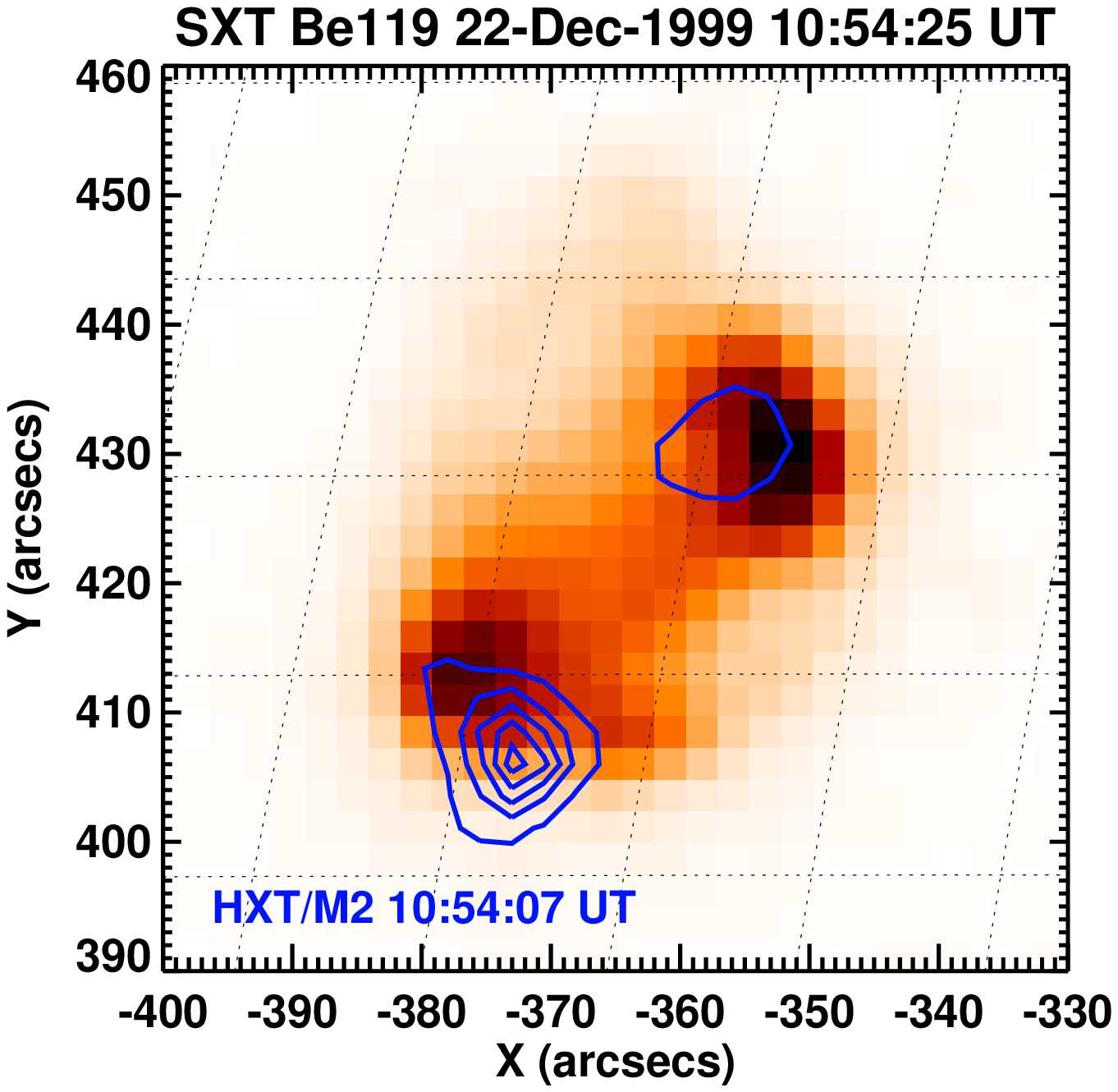}
\includegraphics[width=5.0cm]{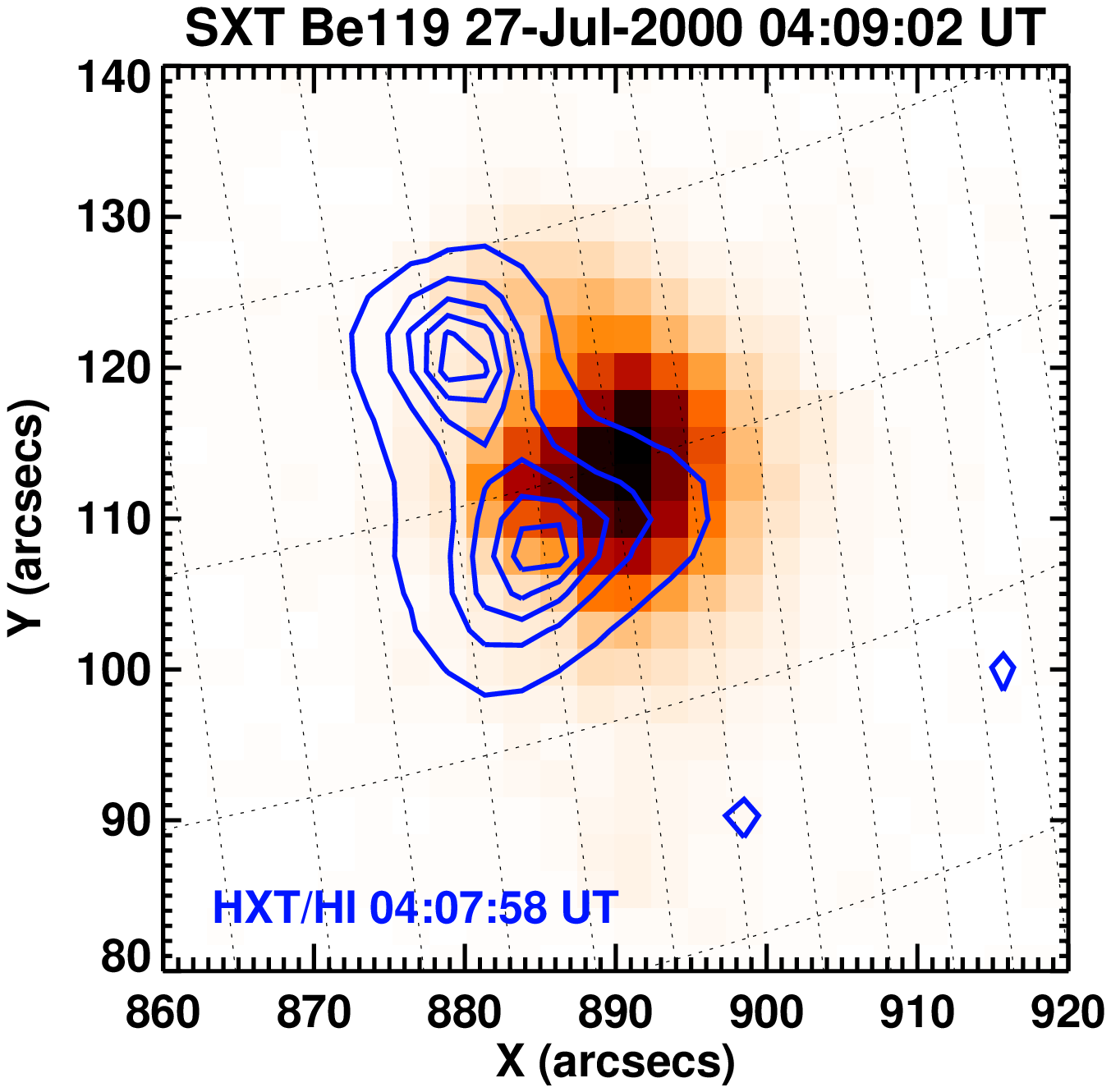}

\includegraphics[width=5.0cm]{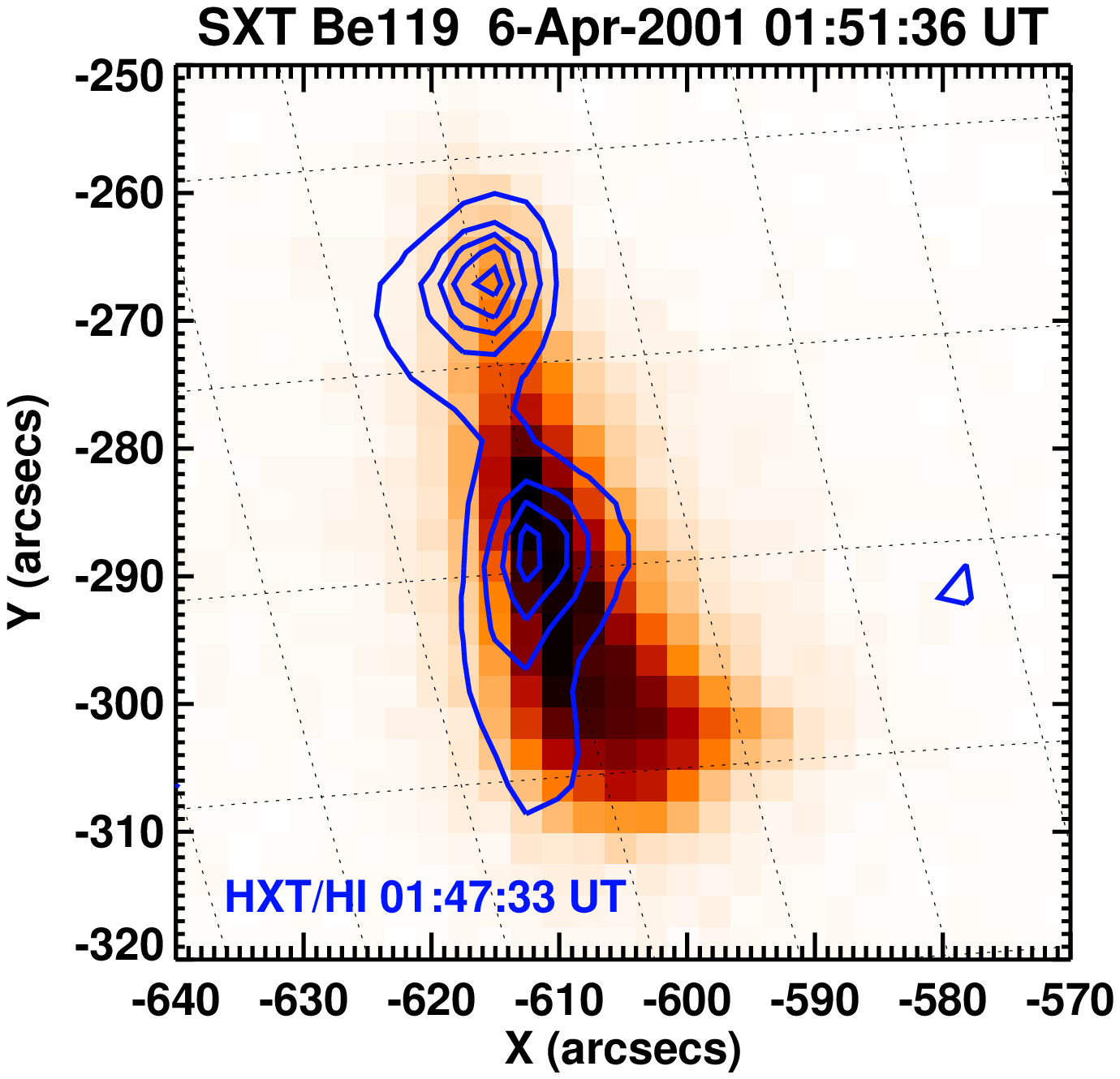}
\includegraphics[width=5.0cm]{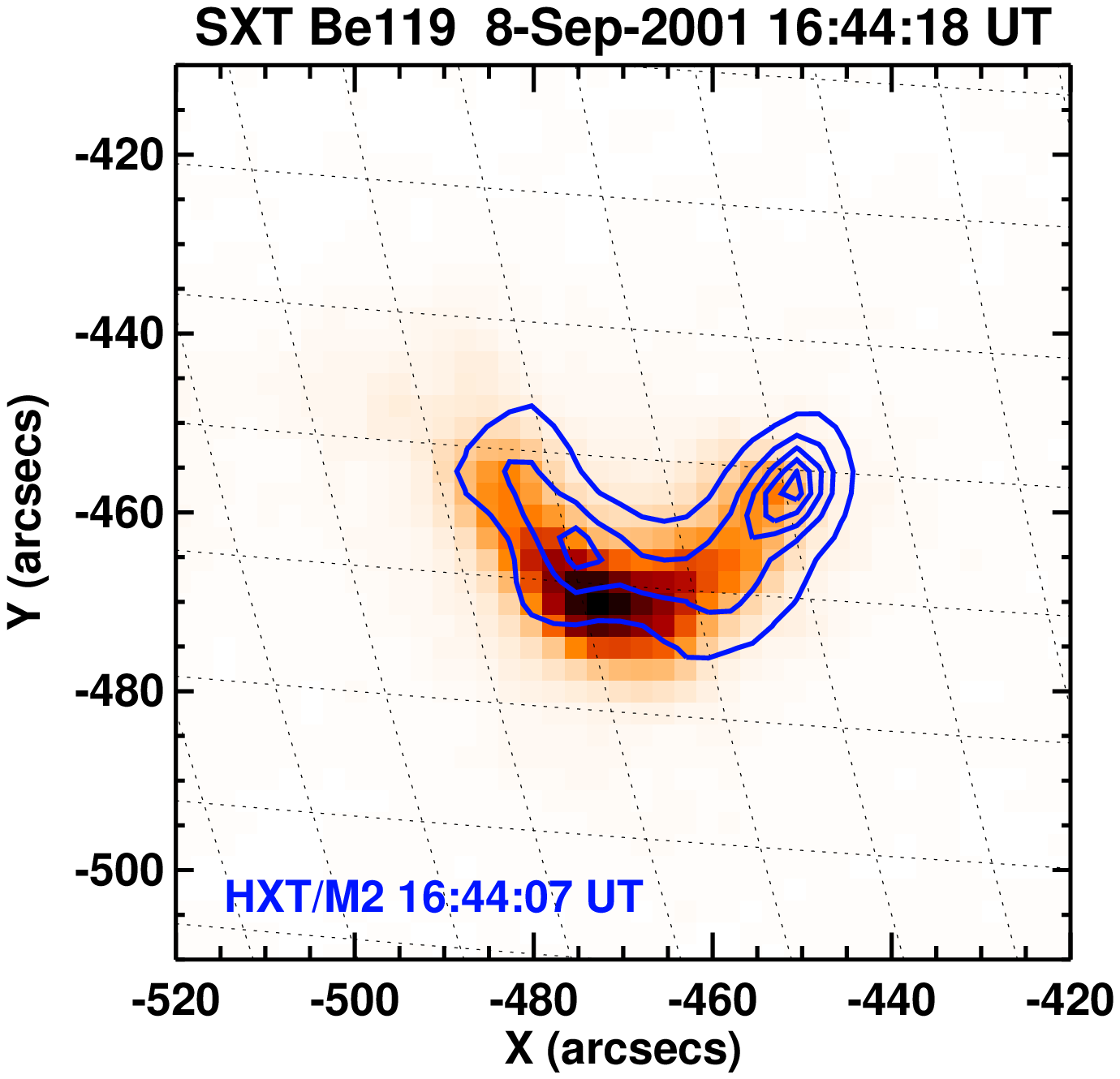}
\includegraphics[width=5.0cm]{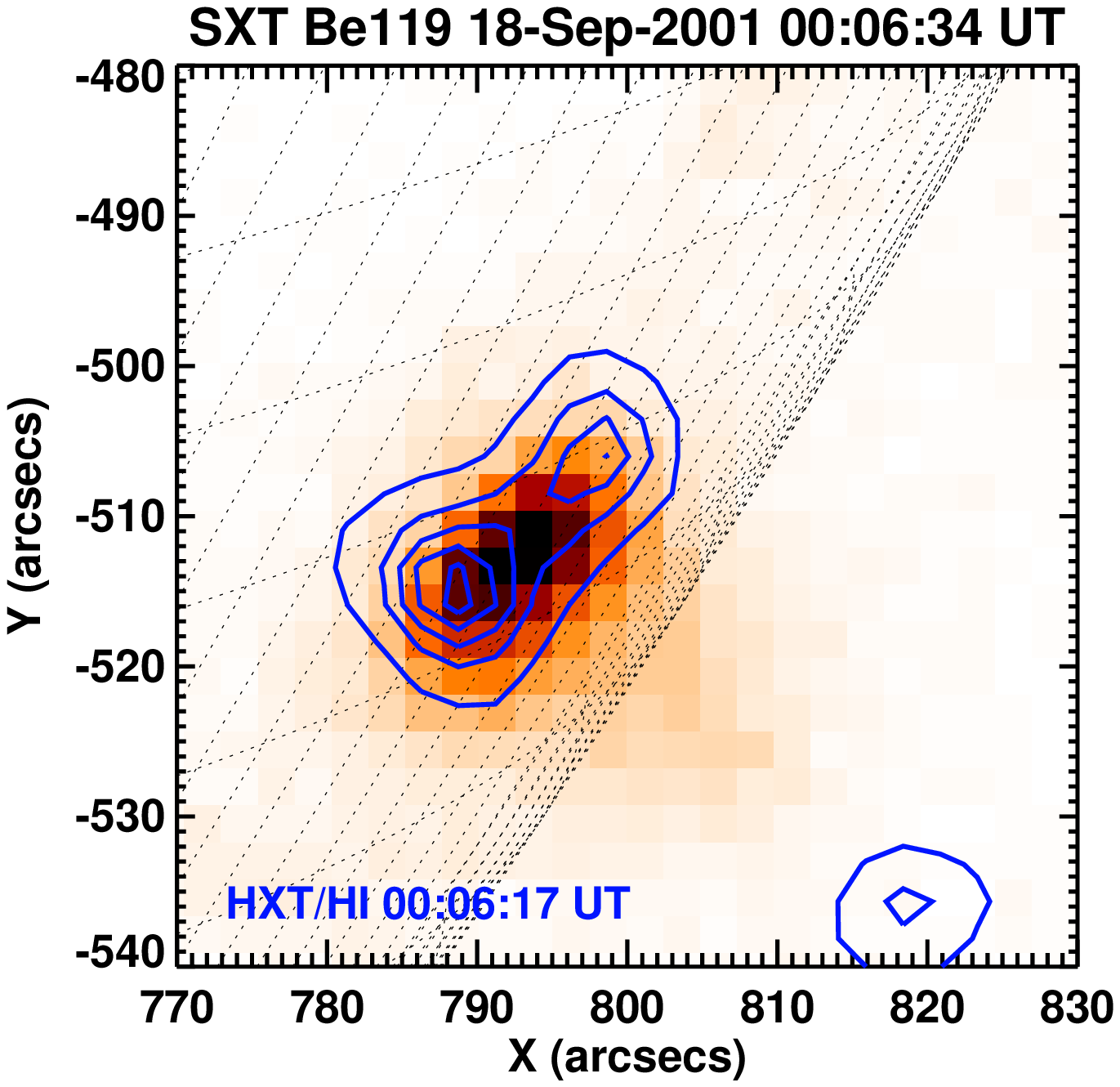}

\caption{Images of twelve analyzed flares taken with the {\it
Yohkoh} SXT or HXT/LO (gray scale images) and HXT/M2 or HXT/HI
(contours) instruments.} \label{fig01}
\end{figure*}

In this paper we investigate the influence of non-thermal electron
energy distribution on the resulting \textit{GOES} classes of flares
and energy used by chromospheric evaporation. Our goal was to
separate the influence the spectral index ($\delta$) and low energy
cut-off ($E_c$) of the non-thermal electron distribution, taken to
be a power law in energy, from the total energy of the electrons. To
do this we calculated a grid of 1D models to describe the time
evolution of twelve observed solar flares. All the models of each
observed flare were calculated using identical total energies
delivered by non-thermal electrons but with various appropriate
combinations of $\delta$ and $E_c$. For each model, we calculated
the resulting \textit{GOES} $1-8$ \AA~flux (i.e. \textit{GOES} class
of the event) as well as the evaporation energy.

In the following, we describe the observed flares (Section 2), the
model calculation (section 3), the results obtained (Section 4) and
the discussion and conclusions are given in Section 5.

\begin{table*}[t!]
\caption{Physical parameters of observed flares.} 
\label{table:1} 
\centering 
\begin{tabular}{c c c c c c c c c c} 
\hline\hline 
Event     & Time of & GOES    & GOES         &Type             &$\gamma$ & $a_0$                & $E_c$ &  S              & $L_0$        \\ %
date      & maximum & class   & incremental  &of               &         &                      &       &                  &              \\
          & [UT]    &         & class        &flare            &         & [ph/$cm^2$/sec/keV]  & [keV] & [$10^{17} cm^2$] & [$10^{8}cm$] \\
 \hline
16-Dec-91 & 04:58   & M2.8    & M2.7         &C                & 3.6    & $7.6\times 10^{6}$ & 25.8   & 3.9             &  15.4       \\
02-Feb-92 & 11:34   & C5.5    & C2.7         &N                & 3.0    & $2.4\times 10^{5}$ & 18.9   & 2.3             &  09.8       \\
27-Oct-92 & 01:47   & M1.1    & C9.5         &C                & 4.5    & $2.8\times 10^{7}$ & 25.2   & 2.3             &  11.6       \\
03-Oct-93 & 09:11   & C1.0    & B8.1         &N                & 2.7    & $6.2\times 10^{4}$ & 23.0   & 1.4             &  02.9       \\
07-Mar-93 & 21:48   & C1.5    & B9.2         &N                & 3.0    & $4.8\times 10^{4}$ & 24.0   & 2.3             &  12.6       \\
26-Jan-94 & 05:41   & C1.4    & C1.1         &N                & 2.6    & $4.9\times 10^{4}$ & 17.2   & 1.3             &  19.5       \\
30-Jun-99 & 11:30   & M1.9    & M1.8         &C                & 3.4    & $1.0\times 10^{6}$ & 25.8   & 1.8             &  11.5       \\
22-Dec-99 & 10:56   & C6.4    & C5.4         &N                & 3.7    & $2.8\times 10^{6}$ & 25.2   & 3.9             &  16.6       \\
27-Jul-00 & 04:10   & M2.5    & M2.4         &C                & 3.2    & $4.8\times 10^{5}$ & 19.8   & 2.3             &  08.7       \\
06-Apr-01 & 01:49   & C7.8    & C4.9         &N                & 2.7    & $2.1\times 10^{5}$ & 28.7   & 2.3             &  13.0       \\
08-Sep-01 & 16:45   & C5.1    & C3.2         &N                & 2.9    & $2.0\times 10^{5}$ & 29.6   & 2.3             &  17.0       \\
18-Sep-01 & 00:08   & M1.5    & M1.3         &C                & 3.7    & $2.3\times 10^{6}$ & 27.0   & 1.1             &  06.8       \\

\hline 
\multicolumn{10}{l}{{\tiny $\gamma$ - photon spectral index; $E_c$ - low energy cut-off; $a_0$ - scaling factor (flux at 1 keV);}}\\
\multicolumn{10}{l}{{\tiny $S$ and $L_0$ - cross-section and semi-length of the flaring loop; Type of flare: $C$ - correlated flare; $N$ - non-correlated flare.}} \\
\vspace{0.01cm}
\end{tabular}
\end{table*}

\section{Observations}

We selected 12 disk flares observed by the \emph{Yohkoh} satellite,
having a simple single-loop X-ray structure and maximum hard X-ray
flux not less than 10 ctns/sec per subcollimator of the M2 channel
(33-53 keV) of the HXT instrument (\citet{Kos91}). Details are given
in Table 1. Five flares were analyzed by McDonald et al. (1999)
(four being correlated and one non-correlated), a further seven
flares were taken from the HXT Flare Catalogue (\citet{Sato06}),
four being correlated and three non-correlated. The correlated and
non-correlated events are denoted in Table 1 with letters C and N
respectively.

The flares were also observed with the \textit{Yohkoh} SXT
grazing-incidence telescope (\citet{Tsu91}) and Bragg Crystal
Spectrometer (\emph{BCS}; \citet{Culhane}) as well as the
\textit{GOES} X-ray photometers (1-8 \AA~and 0.5-4 \AA~bands). HXT
images of the flares were reconstructed using a standard Pixon
method (\citet{Met96}) with variable accumulation times and an
assumed threshold count rate of 200 counts in the M2 band (33-53
keV); these are shown in Figure 1. SXR images (also shown) of ten
flares were taken with Be119/SXT, but for two flares, due to a lack
of the SXT images, we present images taken with HXT in its L0 (14 -
23 keV) channel.

HXT spectra were analyzed to give the photon spectral index
($\gamma$) at the flare peak time in the M2 channel, flux scaling
factor (flux at 1 keV = $a_0$) and cut-off energy in the electron
distribution ($E_c$). The SXT images allowed us to estimate the
single loop semi-length ($L_0$) and cross-section ($S$). These are
given in Table 1.

\begin{figure*}[t!]
\centering \vspace{0.5 cm}
\includegraphics[width=6.8cm]{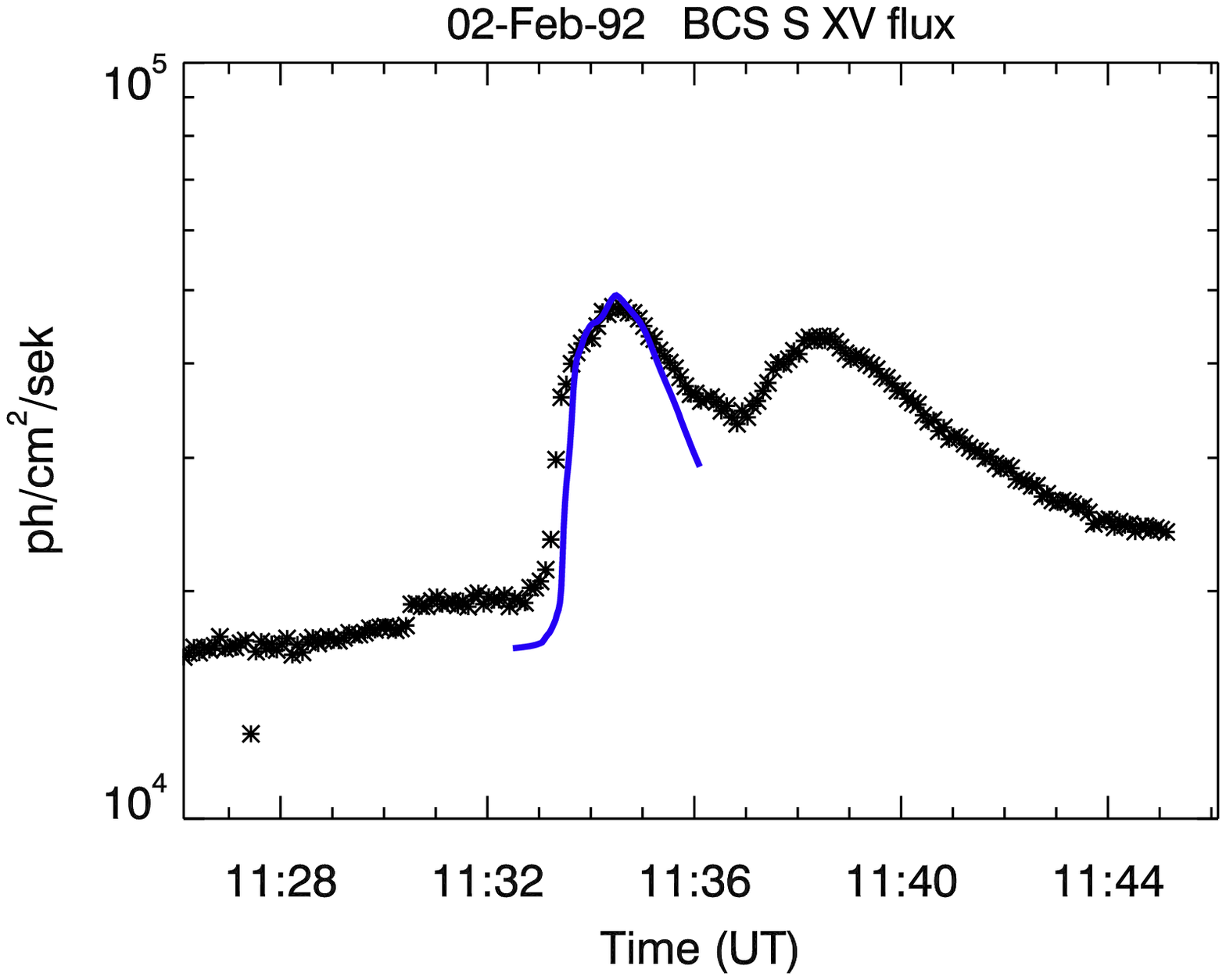}
\includegraphics[width=6.8cm]{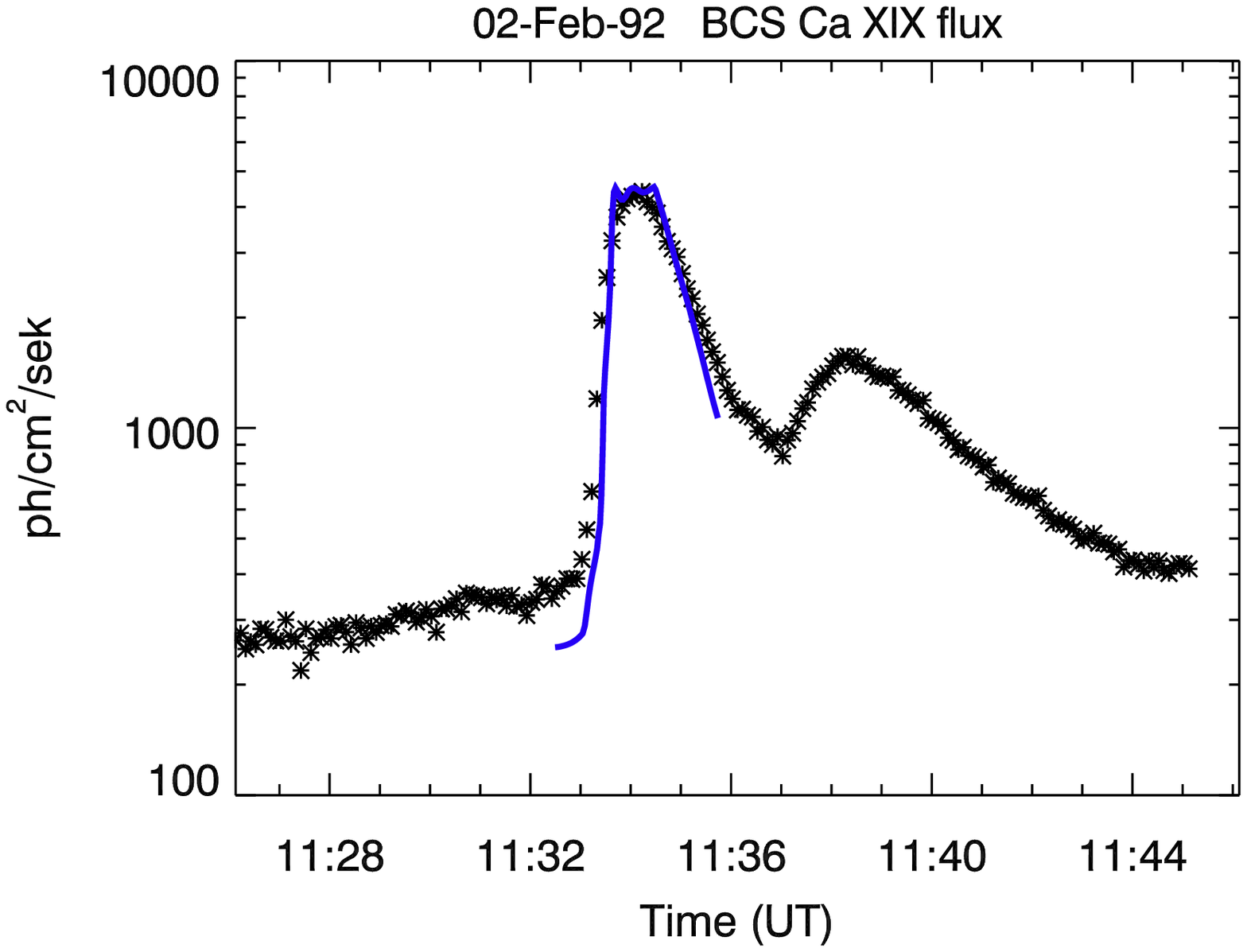}
\vspace{0.5 cm} \caption{ Synthesised BCS CaXIX and BCS SXV fluxes
(blue lines), and the observed fluxes (asterisks) of the C2.7 class
solar flare observed on 1992 February 2. } \label{fig02}
\end{figure*}

\section{Methods of analysis}

Solar flare hard X-ray emission $\gtrsim$ 10-20 keV is generally
believed to be produced by bremsstrahlung emitted by the non-thermal
electrons. A power-law dependence of the emitted spectrum implies a
power law energy distribution of the non-thermal electrons
(\citet{Brown1971}; \citet{Tandberg}) and this in turn requires a
low energy cut-off $E_c$ of the electron spectrum to prevent an
infinite total electron energy. A value of $E_c$ has been arbitrary
set by numerous authors between 15 and 25 keV (see e.g. McDonald et
al. (1999) and references therein). Thus $E_c$ is in the energy
range where thermal emission can dominate and so it is very
difficult to determine its correct value from observations.

With the improved spectral resolution of the \emph{RHESSI}
instrument, the thermal and non-thermal spectral components can be
more clearly separated. Also, direct inversion of photon spectra is
now possible to deduce the "mean electron flux distribution"
(\citet{Brown2006}). But the problem of estimation $E_c$ remains
(e.g. \citet{Kontar2008}). In the absence of \emph{RHESSI} data for
these flares our approach has been to compare the \emph{Yohkoh}/HXT
observations of the 12 flares with results of model calculations. We
investigated the relation between the non-thermal electron spectra
and the \textit{GOES} class of the thermal flare which is produced
by evaporation processes using observed parameters of the 12 flares.
Our calculations took into account the observed energy distributions
of the non-thermal electrons and time variations of the observed
X-ray fluxes, dimensions of the flaring loops from SXR images,
estimated main initial physical parameters of the thermal plasma
(density, temperature), and energy gain and losses.

We analysed each flare in two steps: (a) we built a model of the
observed flare that most closely resembled the synthesized and
observed \emph{GOES} and BCS light curves (when available); (b) we
investigated how the variations of the non-thermal electron energy
spectra influence the synthesized \textit{GOES} fluxes of solar
flare model.

The geometry of each flaring loop (volume ($V$), loop cross section
($S$) and half-length ($L_0$)) were determined using images taken
with SXT and HXT (see Table 1). The loop cross sections were
estimated as the areas within a level equal to 30\% of the maximum
flux in the HXT/M2 channel. Loop half lengths $L_0$ were estimated
from the distances between the centres of gravity of the HXT/M2
footpoints, assuming a semi-circular shape for the loop. The volume
of the loop $V$ then equals $2\,L_0\, S$. Temperatures ($T_e$) and
emission measures ($EM$) were estimated using \textit{GOES} 1 - 8
\AA~ and 0.5 - 4 \AA~ fluxes using the filter-ratio method proposed
by Thomas et al. (1985). We used here an updated version of this
paper by White et al. (2005). A detailed description of this method
is given also by Siarkowski et al. (2008). Mean electron densities
($n_e$) were estimated from emission measures (EM) and volumes ($V=2
L_{0} S$).

\begin{figure}[t]
\centering
\includegraphics[width=8.5cm]{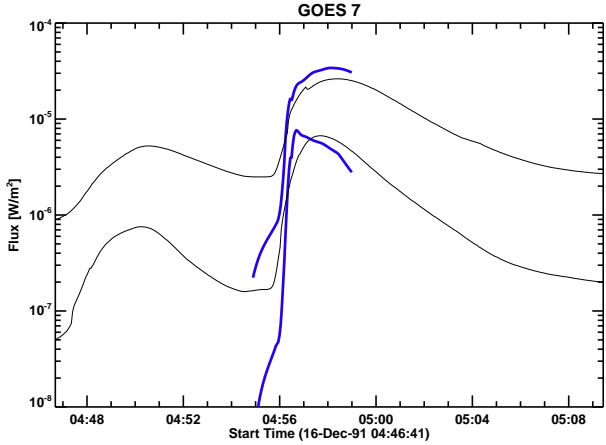}
\vspace{0.5cm} \caption{ Synthetic X-ray fluxes in 0.5-4 \AA~ and
1-8 \AA~ bands calculated using a numerical model of the M2.7 class
solar flare observed on 1991 December 16 (blue thick line)  and the
corresponding fluxes recorded with the \textit{GOES} 7 satellite
(thin grey lines). } \label{fig03}
\end{figure}

Assuming a power-law hard X-ray photon spectra $I(\epsilon) =
a_{0}\epsilon^{-\gamma}$ we calculated time variations of the
spectral indexes ($\gamma$) and scaling factors $a_0$ (flux at 1
keV) for the impulsive phases of all 12 flares. Electron spectra of
the form of $F= A\,E^{-\delta}$ can be calculated from power-law
photon spectra using thick target approximation (\citet{Tandberg}):

\begin{equation}
\label{ } A=a_0 \, \frac{4\pi R^{2}C}{S\kappa_{BH}\bar{Z^{2}}}
\frac{\gamma(\gamma-1)}{B(\gamma-1, \frac{1}{2})}\;\;\;{\rm and}
\;\;\; \delta=\gamma+1
\end{equation}
where $S$ is the loop cross-section area, \textit{R} is the
Earth-Sun distance (1 AU $\simeq 1.50 \times 10^{13}$ cm),
$\;\kappa_{BH}$ is the constant in the Bethe-Heitler cross-section
($7.9 \times 10^{-25} {\rm cm^{2}}\;{\rm keV}$), and $\;\bar{Z^{2}}$
is the abundance-weighted value of an atomic mass in the solar
atmosphere, assumed to be equal to 1.4. The parameter C is defined
as $2\pi e^4 \Lambda \approx 2.6 \times10^{-18}\, {\rm cm^{2}
keV^{2}}$, where $\Lambda$ is the Coulomb logarithm, and $B(x,y)$ is
the complete beta function. The total energies radiated in the SXR
range were calculated by summing energies emitted by the flaring
loop in all time-steps of the model over the duration of the flare
impulsive phase, using for each time-step the measured temperature
($T_{e}$), electron density ($n_{e}$) and corresponding emission
function.

The estimated total energy carried by the non-thermal electrons is
very sensitive to the assumed low-energy cut-off of the electron
spectrum, $E_c$, due to the power-law nature of the energy
distribution. A change of the $E_c$ value by just a few keV can add
or remove a substantial amount of energy from/to the modeled system,
so $E_c$ must be selected with great care.We estimated $E_c$ as
follows. First, we derived low energy cut-offs of the energy spectra
using the slightly modified semi-analytical model of McDonald et al.
(1999). While these authors used a fixed value of $E_c$ (equal to 20
keV) for all flares, we calculated $E_c$ for each flare using an
iterative method. The flare energy budged was calculated using
\textit{GOES} and HXT data. When the total energies calculated from
HXR data and those emitted in the SXR range disagrees, we
iteratively changed the value of $E_c$ (in steps of 0.1 keV) until
agreement was achieved. This method is the most direct way to
estimate $E_c$ from the energy balance.We then slightly modified the
value of $E_c$ further in order to obtain the best agreement of the
synthesised and observed \textit{GOES} classes and BCS fluxes for
each flare. Examples of the agreement between observed and
synthesised BCS light curves are shown in Figure 2 while the
observed and synthesised \emph{GOES} curves are shown in Figure 3.
The estimated $E_c$ was used as a fixed value while electron
spectral index $\delta$ and corresponding scaling factor $A$ varied
in time for each particular model of event. The values of the main
physical parameters, including $\gamma$ and estimated $E_c$,
observed at maximum of the HXR emission, are given in Table 1 for
all twelve flares.

By changing the electron energy spectral index $\delta$ and
appropriate adjustment of $A$ and $E_c$, one can easily change the
amount of the energy used for the evaporation process but keeping
fixed the total energy flux delivered by non-thermal
electrons\textbf{:}

\begin{equation}
\label{ } \Phi = \int_{\;E_c}^{\;\infty} AE^{-\delta}E\,dE =
\frac{A}{\delta-2}\,E_{c}^{2-\delta}
\end{equation}

For each time step during each flare and with values of $\delta$ and
$E_c$ chosen such that $\Phi$ is constant, one can calculate an
appropriate value of $A$ from (2). This allows one to generate a set
of possible electron beams, all with identical total energy
(integrated over the whole impulsive phase of the event) and
identical time variations of the energy flux, and to use them to
calculate a grid of models for each flare (see Fig. 4). For this
work, we defined the impulsive phase of each flare as the period
when the hard X-ray flux as recorded in HXT/M1 channel was larger
than 10 percent of the maximum flux.

\begin{table*}[ht!]
\caption{Results of the numerical models of analysed flares} 
\label{table:1} 
\centering 
\begin{tabular}{c c c c c c} 
\hline\hline 
          & OBSERVATIONS      & \multicolumn{4}{c}{MODEL}                                     \\\cline{3-6}
Event     & GOES              &\multicolumn{2}{c}{Minimum}   & \multicolumn{2}{c}{Maximum}    \\ %
date      & class$^*$         &GOES      &$E_{evap}/E_{nth}$ &GOES       &$E_{evap}/E_{nth}$  \\
          &                   &class$^*$ &                   &class$^*$  &                    \\
 \hline
16-Dec-91 & M2.8 (M2.7)       &C2.4 (C1.4)& 0.07              &M9.5 (M9.4)& 0.79               \\
02-Feb-92 & C5.5 (C2.7)       &C3.4 (B6.3)& 0.17              &C8.4 (C5.6)& 0.78               \\
27-Oct-92 & M1.1 (C9.5)       &C2.5 (B9.5)& 0.08              &M1.8 (M1.6)& 0.82               \\
03-Oct-93 & C1.0 (B8.1)       &B6.5 (B4.4)& 0.19              &C2.8 (C2.6)& 0.71               \\
07-Mar-93 & C1.5 (B9.2)       &B9.5 (B3.7)& 0.07              &C2.4 (C1.8)& 0.70               \\
26-Jan-94 & C1.4 (C1.1)       &B7.4 (B4.4)& 0.22              &C2.8 (C2.5)& 0.84               \\
30-Jun-99 & M1.9 (M1.8)       &C2.0 (B9.9)& 0.04              &X1.1 (X1.1)& 0.71               \\
22-Dec-99 & C6.4 (C5.4)       &C2.1 (C1.1)& 0.06              &M4.5 (M4.4)& 0.66               \\
27-Jul-00 & M2.4 (M2.3)       &C4.2 (C3.2)& 0.11              &M4.7 (M4.6)& 0.80               \\
06-Apr-01 & C7.8 (C4.9)       &C4.5 (C1.5)& 0.07              &M3.1 (M2.8)& 0.68               \\
08-Sep-01 & C5.1 (C3.2)       &C1.9 (A2.4)& 0.02              &M1.4 (M1.2)& 0.51               \\
18-Sep-01 & M1.5 (M1.3)       &C3.1 (C1.1)& 0.05              &M4.8 (M4.6)& 0.72               \\
\hline 
\multicolumn{6}{l}{{\tiny $^*$\textit{GOES} class minus the preflare level is given in parentheses.}} \\
\vspace{0.01cm}
\end{tabular}
\end{table*}

In this work we used a modified Naval Research Laboratory Solar Flux
Tube Model code kindly made available to the solar community by
Mariska and his co-workers (Mariska et al. (1982), Mariska et al.
(1989)). Although a typical flaring loop is a 3-D structure
surrounded by a possibly complex active region, it can be modeled
for many purposes with a simple 1-D hydrodynamic model as with the
NRL Code. We included a few modifications to the code: new radiative
loss and heating functions; inclusion of the VAL-C model
(\citet{Vernaz}) of the initial structure of the lower part of the
loop (extended down using Solar Standard Model data
(\citet{bahcall}); and use of double precision in the calculations.
The heating of the plasma by a non-thermal electron beam was modeled
using the approximation given by Fisher (1989). A mesh of new values
of the radiative loss function was calculated using the CHIANTI
(version 5.2) software (\citet{Dere}, \citet{Landi}) in a
temperature range $10^4 - 10^8$ K and density range $10^8 - 10^{14}$
${\rm cm^{-3}}$. For each flare a grid of models was calculated
using various non-thermal electron beam models, all having the same
total energy. All models were calculated for periods lasting from
the beginning of the impulsive phase to beyond the soft X-ray
emissions maximum. These periods were about 150-200 seconds. The
time steps in the models were about 0.0005-0.001 sec.

We estimated the evaporation energy $E_{evap}$ as the difference
between the total energy delivered by the non-thermal electrons
$E_{nth}$ and the total energy lost by radiation over the whole
loop, $E_{rad}$~ (i.e. $E_{evap}=E_{nth}-E_{rad}$) over the
impulsive phase. As mentioned before, the main parameters of each
calculated model (time-dependent fluxes of the non-thermal electrons
and the lengths and cross-sections of the flaring loops) were
evaluated using observational data (see Section 2).

\section{Results}

We analysed each flare in two steps. First, we built a model
 using observed geometrical and physical parameters, including $\delta$
 and $E_c$. The correctness of these models was
validated by comparison of the synthesized \textit{GOES} 1-8 \AA~and
0.5-4 \AA, BCS SXV and BCS CaXIX fluxes with the observed fluxes. We
achieved good agreement between observations and the modeled data.
For example, the observed and synthesized BCS SXV and BCS CaXIX
fluxes of the C2.7 class solar flare observed on 1992 February 2 are
shown in Figure 2 while the calculated and observed \textit{GOES}
X-ray light-curves of the M2.7 event on 1991 December 16 are shown
in Figure 3. The values of the flux maxima are reproduced very well
but the large discrepancies noticeable during the early rise and
late decay phases are caused by lack of an additional pre- and
post-impulsive flare heating in our model. In the second step, we
calculated several numerical models of each event for various
appropriate combinations of the electron spectral index $\delta$ and
low-energy cut-off $E_c$, keeping fixed the total energy delivered
by the non-thermal electrons equal to the observed energy.

The model calculations also provide pressure, temperature, density,
velocity, column mass etc. as function of time and position along
the loop. For all reasonable sets of physical parameters of the
flaring loops we found a fast inflow of chromospheric material into
the loop (i.e. chromospheric evaporation). Large-scale macroscopic
motions of the dense plasma toward the loop-tops and fast increases
of the plasma temperature, pressure, electron density all agree well
with commonly accepted schemes of chromospheric evaporation. After
the energy deposition period, the plasma contained in the flaring
loop gradually cools but the model calculations generally ended
before returning to a hydrostatic equilibrium.

For all twelve flares we found that keeping total energy delivered
by non-thermal electrons fixed, the resulting observed \textit{GOES}
class of the induced solar flare varied significantly when the
spectral index and low energy cut-off of the non-thermal electrons
spectra were changed. The variations of the \emph{GOES} classes and
the proportion of the total energy contributed by
 evaporation are given in Figure 4. The upper left
panel shows models of M2.7 \emph{GOES} class correlated flare
observed on 1991 December 16. The observed low energy cut-off $E_c$
is equal to 25.8 keV and is assumed constant during the
calculations. The spectral index $\delta$ is equal to 4.6 (here and
for the other three events shown on Figure 4 we give the value of
$\delta$  at the time of maximum of the impulsive phase). The upper
right panel shows models of the C2.7 \emph{GOES} class
non-correlated flare observed on 1992 February 2. The observed $E_c$
and $\delta$ are equal to 18.9 keV and 4.0. The lower left panel
shows models of the B8.1 \emph{GOES} class non-correlated flare
observed on 1993 October 3. Observed $E_c$ and $\delta$ are equal to
23 keV and 3.7. Bottom right panel shows models of the M2.4
\emph{GOES} class correlated flare observed on 2000 July 27.
Observed $E_c$ and $\delta$ are equal to 19.8 keV and 4.2. The
spectral indexes $\delta$ of each model varied in time in accordance
with its observational values increased or decreased by a constant
factor. In other words, in our models, we took various values of
$\delta$ that were different than $\delta_{obs}$ by amounts from 0
up to $\pm$2, where $\delta_{obs}$ is the observed $\delta$ at any
given time during the flare. The black, filled squares represent
models calculated using non-thermal electron beams having main
parameters deduced from observations. The total ranges of the
modeled \emph{GOES} classes and $E_{evap}$ for all flares are shown
in Table 2.

We now describe results obtained for the most representative events,
two correlated flares (1991 December 16, 2000 July 27) and two
non-correlated flares (1992 February 2 and 1993 October 3). All
flares had a simple, single-loop structure. (In the following GOES
classes are given with the pre-flare level subtracted.)

\subsection{M2.7 flare on 1991 December 16}

The M2.7 flare at 04:54 UT on 1991 December 16 occurred in active
region NOAA 6961 (N04W45). The flare appeared in SXR as a single
loop with semi-length 15 400 km (see Fig. 1). The impulsive HXR
($>$23 keV) occurred between 04:55:50 UT and 04:56:28 UT. The
cross-section of the loop was estimated from HXR images to be
$S=3.9\times 10^{17}$~cm$^{2}$.

\begin{figure*}[ht!]
\centering
\includegraphics[width=7.5 cm]{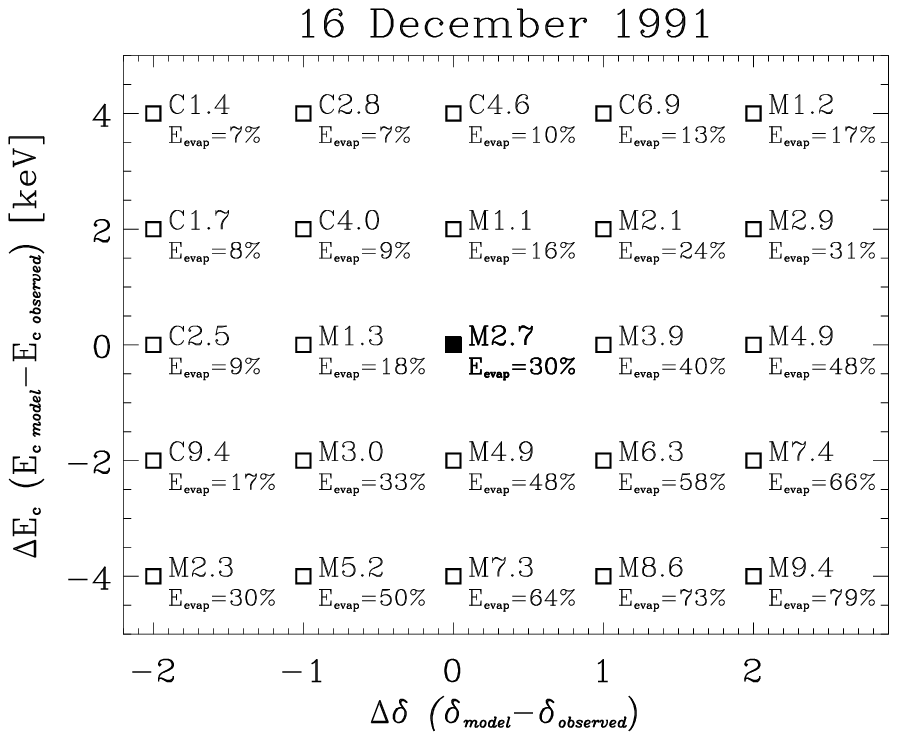}
\includegraphics[width=7.5 cm]{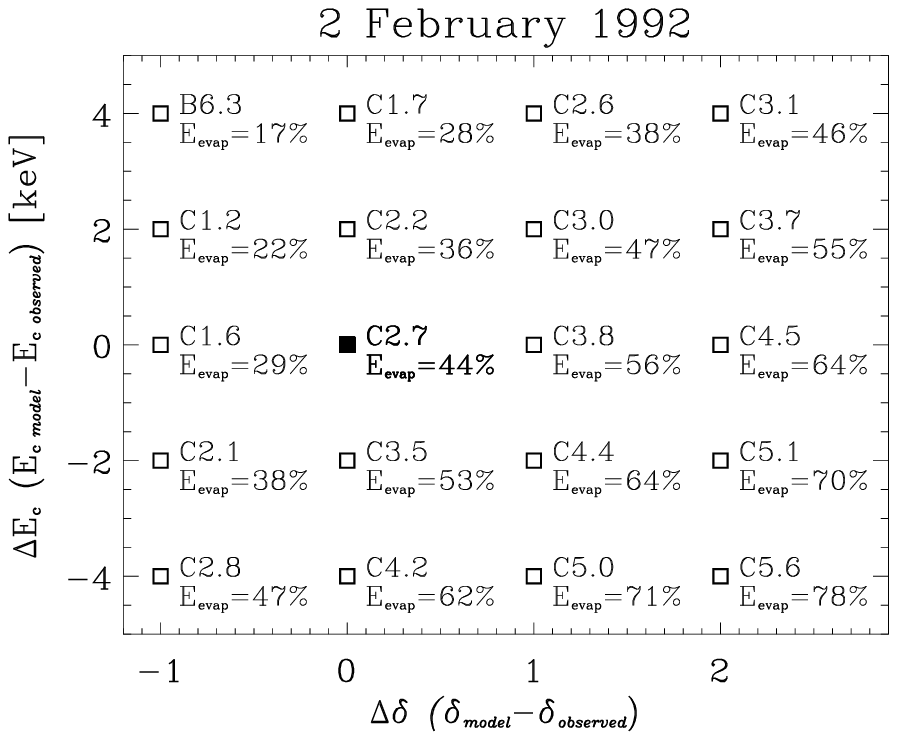}

\includegraphics[width=7.5 cm]{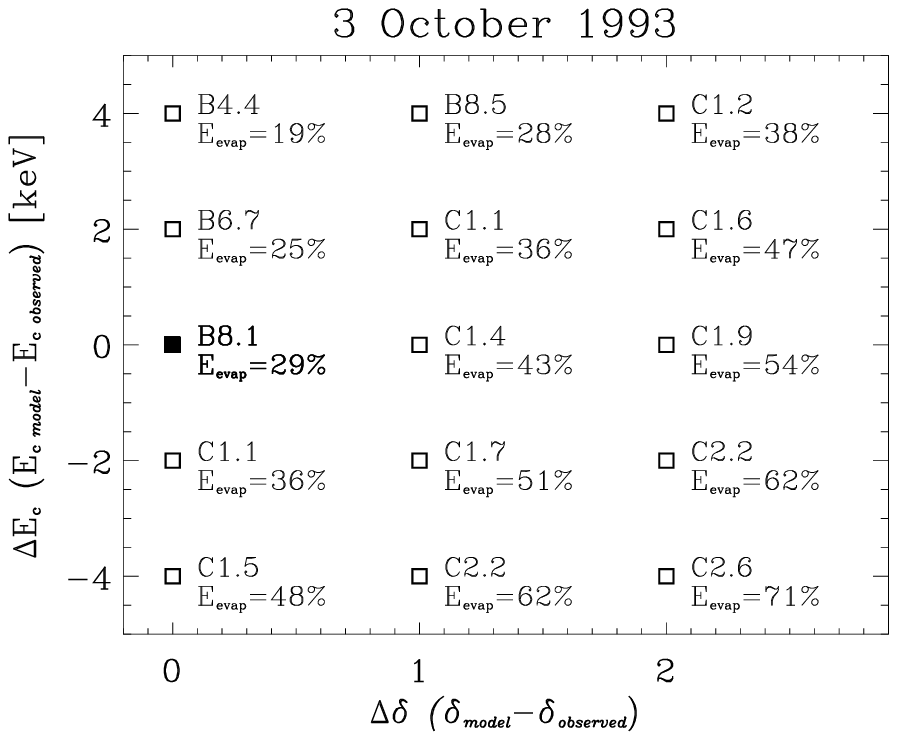}
\includegraphics[width=7.5 cm]{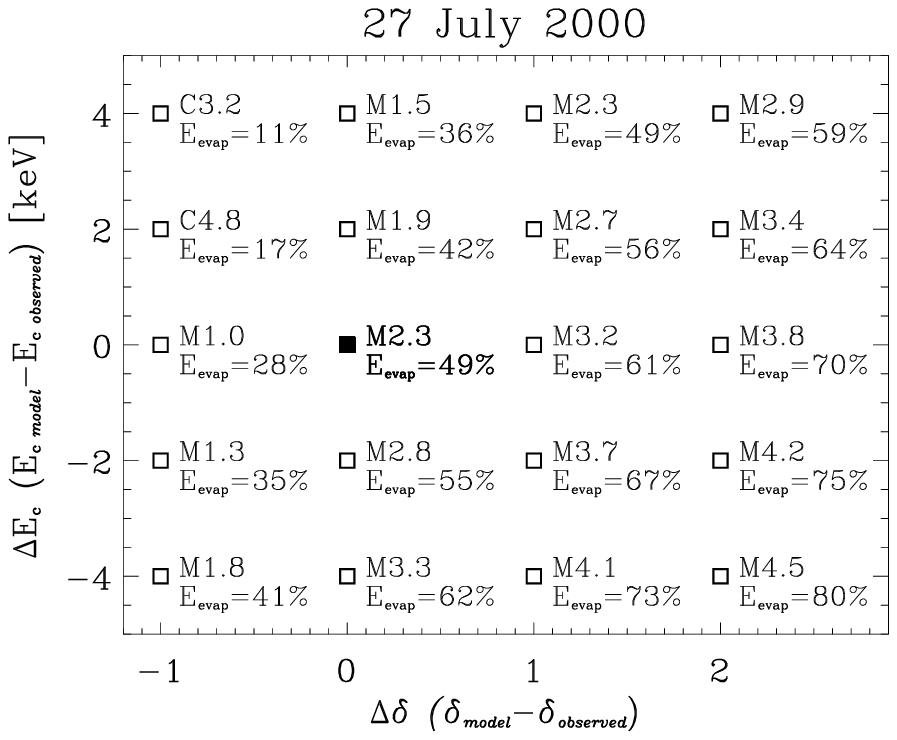}

\caption{Variations of the observed \emph{GOES} classes and shares
of the evaporation energy $E_{evap}$ in the total energies delivered
by non-thermal electrons for four solar flares observed on 1991
December 16, 1992 February 2, 1993 October 3 and on 2000 July 27.
See Section 4 for details.} \label{fig04a}
\end{figure*}

A flare model was calculated using parameters of the photon
spectra $\gamma$ and scaling factors $a_{0}$ calculated as a
function of time from hard X-ray fluxes observed in the HXT $M1$
and $M2$ channels, while $E_{c}$ was estimated and fixed as equal
to 25.8 keV. The total non-thermal electron energy estimate to be
$E_{nth}=1.2\times 10^{30}$ ergs while the evaporation energy was
 $E_{evap}=3.6\times 10^{29}$ ergs ($=0.3\, E_{nth}$). Alternative
models were calculated for the same total energy $E_{nth}$ but with
$\delta$ changed from the initial value of 4.6 by $\pm 2$ and with
$E_c$ changed from the initial value of $E_c$ by $\pm 4$ keV (see
Figure 4, top left panel). For $\delta$ lowered by 2 and $E_c$
raised by 4 keV the evaporation energy decreased to 7\% of the
$E_{nth}$ only while the \textit{GOES} class decreased to only C1.4.
By contrast, for $\delta$ increased by 2 and $E_c$ lowered by 4 keV
the observed \textit{GOES} class of the event increased to M9.4
while the evaporation energy increased to 79\% of $E_{nth}$.
Calculations made for all combinations of $\delta$ and $E_c$ give
flares of with \textit{GOES} class varying between C1.4 and M9.4.

\subsection{C2.7 flare on 1992 February 2}

The C2.7 flare on 1992 February 2 at 11:33 UT - 11:38 UT occurred in
active region NOAA 7042 (S14W41). It was observed with
HXT/\textit{Yohkoh} and \textit{GOES} only (see Fig. 1). The
impulsive HXR ($>$23 keV) peak occurred between 11:33:18 UT and
11:33:28 UT, only 10~s long. HXT/LO images of the flare show the
X-ray emission as a single loop with semi-length ($\simeq 9 800$
km). The loop foot points were observed in HXT channels M2 and H
only. The loop cross-section was estimated as $S=2.3\times
10^{17}$~cm$^{2}$.

The flare model was calculated as for the previous flare. The total
 energy was $E_{nth}=1.7\times 10^{29}$ ergs and the
evaporation energy was $E_{evap}=7.5\times 10^{28}$ ergs, ($=0.44
E_{nth}$). Alternative models were calculated for the same total
non-thermal electron energy $E_{nth}$ but with $\delta$ in the range
 -1 to +2 from the initial value of $\delta$ and with $E_c$ in the range $\pm4$ keV from the
initial value of $E_c$ (see Figure 4, top right panel). Calculations
made for all combinations of spectral index and low energy cut-off
give flares of \textit{GOES} classes in range B6.3 and C5.6.

\subsection{B8.0 flare on 1993 October 3}

The B8.1 flare on 1993 October 3 occurred between 09:06 UT and 09:07
UT in active region NOAA 7590 (N11W04). The flare was visible in SXT
as a single loop of semi-length $\sim$ 28 800 km (see Fig. 1). The
impulsive HXR ($>$23 keV) emission occurred between 09:06:40 UT and
09:07:15 UT. The cross-section of the loop was estimated as
$S=1.4\times 10^{17}$~cm$^{2}$.

A flare model was calculated using the total energy equal to
$E_{nth}=1.6\times 10^{29}$ ergs. The evaporation energy was equal
to $E_{evap}=4.6\times 10^{28}$ ergs, ($=0.29 E_{nth}$). Alternative
models were calculated for the same total energy $E_{nth}$ but with
$\delta$ changed by 0 to +2 from the initial value of $\delta$ and
with $E_c$ changed from the initial value of $E_c$ by $\pm4$ keV (
see Figure 4, bottom left panel). For the spectral index increased
by 2 and energy cut-off lowered by 2 keV the evaporation energy of
the flare increased to $0.71\,E_{nth}$ and the \textit{GOES} class
of the flare increased to C2.6. Calculations made for all
combinations of spectral index and energy cut-off give flares of
 \textit{GOES} classes varying from B4.4 to C2.6.

\subsection{M2.4 flare on 2000 July 27}

The M2.4 flare on 200 July 27 occurred between 04:08 UT - 04:13 UT
in active region NOAA 9090 (N10W72). The flare was visible in SXT as
a single loop with semi-length $\sim 8 700$ km (see Fig. 1). The HXR
($>$23 keV) emission occurred between 04:07:54 UT and 04:08:30 UT.
Using the reconstructed HXR images we estimated the cross-section of
the loop to be $S=2.3\times 10^{17}$~cm$^{2}$.

The flare model was calculated in the same way as for the previous
events. The total energy was equal to $E_{nth}=4.8\times 10^{29}$
ergs and the evaporation energy $E_{evap}=2.3\times 10^{29}$ ergs (
$= 0.44 E_{nth}$). The alternative models of the flare were
calculated for the same total energy $E_{nth}$ but with $\delta$
changed from the initial value of $\delta$ by $-1$ to $-2$ and with
$E_c$ changed from the initial value of $E_c$ by $\pm 4$ keV (see
Figure 4, bottom right panel). For the spectral index increased by 2
and energy cut-off lowered by 4 keV the evaporation energy of the
flare increased to 80\% of the $E_{nth}$ and the \textit{GOES} class
of the flare increased to M4.5. By contrast, for $\delta$ lowered by
1 and $E_c$ raised by 4 keV the observed \textit{GOES} class of the
event lowered to only C3.2 while the evaporation energy decreased to
$0.11\, E_{nth}$. Calculations made for all combinations of $\delta$
and $E_c$ give flares with \textit{GOES} classes varying from C3.2
to M4.6.

\section{Discussion and conclusions}

\begin{figure*}[ht!]
\centering
\includegraphics[width=14 cm]{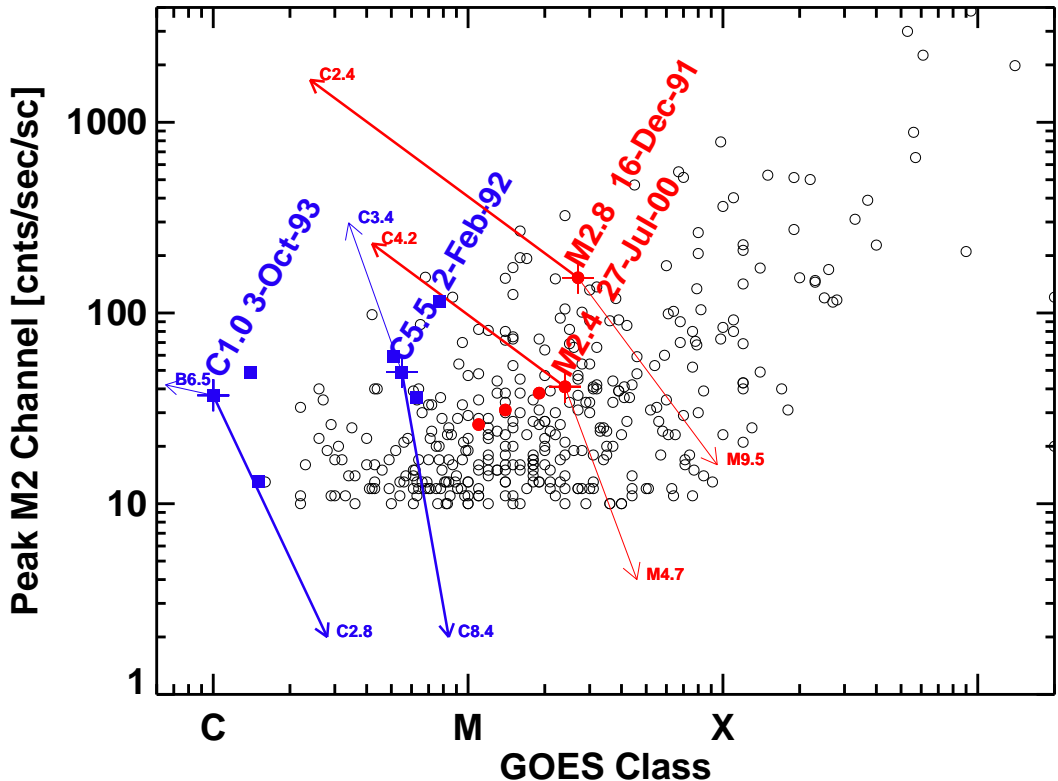}

\caption{Variations of the synthesised \textit{GOES} classes and
X-ray fluxes of the numerical models of the four solar flares
obtained for various $E_c$ and $\delta$ and constant total energy
delivered by non-thermal electrons (see main text and Fig. 4 for
details). The models calculated using observed energy spectra of
non-thermal electrons are marked with crosses. The arrows show the
variations of the \textit{GOES} classes and X-ray fluxes of the
models, the open circles represent 369 flares taken from the Sato
HXT Flare Catalogue. Filled circles (red colour) denote five
correlated  and filled squares (blue colour) seven non-correlated
analyzed flares.} \label{fig06}
\end{figure*}

We have presented results of an energy model calculation for twelve
flares having various soft to hard X-ray emission ratios. Our
interest was focused on the influence of the variations of the
energy spectrum of the non-thermal electrons on the resulting
observed \textit{GOES} classes of the flares and on their
evaporation energies $E_{evap}$. Many numerical models of the events
were calculated using observed and modified energy distributions of
the non-thermal electrons (various appropriate combinations of the
electron spectral index $\delta$ and low energy cut-off $E_c$ but
fixed value of the total energy delivered by non-thermal electrons
$E_{nth}$), observed geometry of the loops and initial values of the
main physical parameters of the plasma estimated using observational
data. We showed first that with fixed initial parameters of the
flare loop (geometry and physical parameters of the plasma) as well
as fixed total non-thermal electron energy, one can significantly
change the resulting \textit{GOES} class of the flare by appropriate
change of the electron index and low energy cut-off of the electron
spectra while maintaining the total energy at a constant value. For
example, we were able to change the M2.7 \textit{GOES} class
correlated flare into only a C1.4 class non-correlated-like flare
and the B8.1 \textit{GOES} class non-correlated flare into a C2.6
class correlated-like flare.

The ratio of the radiated energy to the energy in evaporation
processes as well as observed \textit{GOES} classes of the events
vary for various combinations of the spectral index and low energy
cut-off. It is obvious that both the total delivered energy and
the energy spectrum of the non-thermal electron beam (described by
spectral index and low energy cut-off) are decisive for the course
and magnitude of the evaporation processes as well as for
determination of the observed soft X-ray emission of the flares
(i.e. their \textit{GOES} class).

An estimation of the energy carried by non-thermal electrons is very
sensitive to an assumed value of the low energy cut-off of the
energy spectrum due to the power law nature of the energy
distribution. A change of the $E_c$ value by only a few keV can add
or remove a substantial amount of energy from/to the modeled system.
The spectral index of the non-thermal electron spectrum is also
crucial, while the low-energy fraction of the electrons heats the
chromospheres most effectively (see also McDonald et al., 1999). For
the same total energy the beam of non-thermal electrons with a
softer spectrum could give rise to a solar flare of greater
\textit{GOES} importance. This is demonstrated in Figure 4, where
variations of the observed \emph{GOES} classes and evaporation
energies $E_{evap}$ are shown for four flares. As an example one can
track the behaviour of the M2.7 flare on 1991 December 16. The model
calculated using a non-thermal electron beam having main parameters
deduced from observations, the \emph{GOES} class is equal to M2.7
and the evaporation energy equals 30\% of the total energy delivered
by the non-thermal electrons. The same observed solar flare, modeled
using non-thermal electron beam having spectral index $\delta$
decreased by 1, has a slightly lower \emph{GOES} class equal to M1.3
and significantly lower evaporation energy equal to 18\% $E_{nth}$.
By increasing additionally the low energy cut-off $E_c$ by 2~keV we
obtained an even lower \emph{GOES} class and $E_{evap}$ of the model
equal to C4.0 and 9\% $E_{nth}$, respectively. By contrast, for a
steeper electron spectrum and increased population of the low energy
electrons (decreased $E_c$), the \emph{GOES} classes and $E_{evap}$
increase. Thus, for the M2.7 flare described above, the non-thermal
electron beam with energy spectral index $\delta$ increased by 1 has
a slightly higher \emph{GOES} class equal to M3.9 and higher
evaporation energy equal to 40\% $E_{nth}$. By decreasing
additionally the low energy cut-off $E_c$ by 2~keV (expanding the
low-energy electrons' population) we obtained even higher
\emph{GOES} class and $E_{evap}$ of the model equal to M6.3 and
58\%, respectively. The ranges of the modeled \emph{GOES} classes
and $E_{evap}$ for all analysed events are given in Table 2.

The overall course of changes of \emph{GOES} classes and $E_{evap}$
is similar for all our analysed events. In general, the SXR emission
increases for softer HXR spectra and decreases for harder spectra.
That is, the steeper electron spectra result in a higher amount of
evaporated material. Such behaviour is induced by spatial variations
of an efficiency of the non-thermal electrons' energy deposition
mechanism. The energy deposition mechanism is most efficient in the
transition region and upper part of the chromosphere, but its
particular spatial distribution depends on a actual distribution of
the column mass of plasma encountered by non-thermal electrons.
However, the magnitude of changes observed for a particular flare
depends on the flare's physical proprieties.

A non-thermal electron beam having a large population of high-energy
electrons (i.e. having a hard spectrum) penetrates deep into the
chromospheric plasma where it deposits most of its energy, from
where energy is efficiently radiated. The remaining part of the
carried energy is deposited in the upper part of the chromosphere
and/or transition region, causing moderate "gentle evaporation". On
the other hand, a non-thermal electron beam having soft spectrum
deposits most of its energy in the upper part of the chromosphere
and/or transition region, where the density is relatively low,
giving rise to "explosive evaporation".

Our results show that the parameters and properties of the solar
flares depend not only on the initial hydrodynamic properties of the
flaring loop and on the total amount of the delivered energy but
also on properties of the primary source of energy and time and
spatial variations of the processes leading to the acceleration of
the electrons. Thus, the level of correlation between the cumulative
time integral of HXR and SXR fluxes depends on the HXR energy range.
Such a conclusion is in accordance with results of the statistical
analysis of flares by Veronig et al. (2002) and microflares by Qiu
et. al. (2004) observed with RHESSI indicating that only half of the
events show a time behaviour consistent with the expectations based
on the Neupert effect. Qiu et al. (2004) showed also that the
correlation resulting from the Neupert effect is greatest in the
photon energy range of 14-20 keV.

In Figure 5 we show variations of the synthesised \textit{GOES}
classes and X-ray fluxes of the numerical models of the four solar
flares obtained for various $E_c$ and $\delta$ and constant total
energy delivered by non-thermal electrons. The arrows show the
variations of the \textit{GOES} classes and X-ray fluxes of the
models while the open circles represent 369 flares taken from the
HXT Flare Catalogue (\citet{Sato06}). As can be seen the
relationships between soft and hard X-ray fluxes obtained from the
models go far beyond the range of the "correlation belt" of soft and
hard X-ray fluxes recorded for the observed solar flares, up to high
HXR fluxes. For example, by changing the hardness of the non-thermal
electron spectrum and low energy cut-off the M2.7 flare observed on
1991 December 16 was converted  into the C2.4 \textit{GOES} class
event having an HXT/M2 peak emission above 1000 cnts/sec/sc (see
Figure 5, thick arrow). Many flares with low GOES class but with
large hard X-ray flux have been observed with \emph{Yohkoh} and
\emph{RHESSI}. However, one does not observe flares with similar
GOES class having very large X-ray flux; the Nature apparently do
not realise extremely "small-hard" flares (i.e. those which would be
located in the upper-left corner at the Figure 5). This than imposes
restrictions on the flare electron spectra and therefore on
acceleration mechanisms.

\begin{acknowledgements}
The authors would like to thank the \emph{Yohkoh} team for excellent
solar data and software. They are also grateful to the anonymous
referee for useful comments and suggestions.

This work was supported by the Polish Ministry of Science and Higher
Education, grant number N203 022 31/2991 and by the European
Community's Seventh Framework Programme (FP7/2007-2013) under grant
agreement ${\rm n}^{\circ}$~218816 (SOTERIA).
\end{acknowledgements}


\end{document}